\pdfoutput=1
\documentclass[11pt]{article}

\usepackage[preprint]{acl}

\usepackage{times}
\usepackage{latexsym}
\usepackage{amsmath} 
\usepackage{multirow}
\usepackage{booktabs}
\usepackage[T1]{fontenc}
\usepackage[utf8]{inputenc}

\usepackage{microtype}

\usepackage{inconsolata}

\usepackage{graphicx}
\usepackage{placeins}

\title{Enhancing Temporal Understanding in Audio Question Answering for Large Audio Language Models}

\author{Arvind Krishna Sridhar\\
  Qualcomm Technologies Inc. \\
  San Diego, CA  \\
  \texttt{arvisrid@qti.qualcomm.com} \\\And
  Yinyi Guo \\
  Qualcomm Technologies Inc. \\
  San Diego, CA  \\
  \texttt{yinyig@qti.qualcomm.com}  \\\And
  Erik Visser \\
  Qualcomm Technologies Inc. \\
  San Diego, CA  \\
  \texttt{evisser@qti.qualcomm.com} \\}

\begin{document}
\maketitle
\begin{abstract}
The Audio Question Answering (AQA) task includes audio event classification, audio captioning, and open-ended reasoning. Recently, AQA has garnered attention due to the advent of Large Audio Language Models (LALMs). Current literature focuses on constructing LALMs by integrating audio encoders with text-only Large Language Models (LLMs) through a projection module. While LALMs excel in general audio understanding, they are limited in temporal reasoning, which may hinder their commercial applications and on-device deployment. This paper addresses these challenges and limitations in audio temporal reasoning. First, we introduce a data augmentation technique for generating reliable audio temporal questions and answers using an LLM. Second, we perform a further fine-tuning of an existing baseline using curriculum learning strategy to specialize in temporal reasoning without compromising performance on fine-tuned tasks. We demonstrate the performance of our model using state-of-the-art LALMs on public audio benchmark datasets. Third, we implement our AQA model on-device locally and investigate its CPU inference for edge applications.
\end{abstract}

\section{Introduction}

\begin{figure*}[t!]
\centering
\begin{minipage}[b]{1\linewidth}
  \centering
  \centerline{
  \includegraphics[width=1.1\textwidth,scale=1.0]{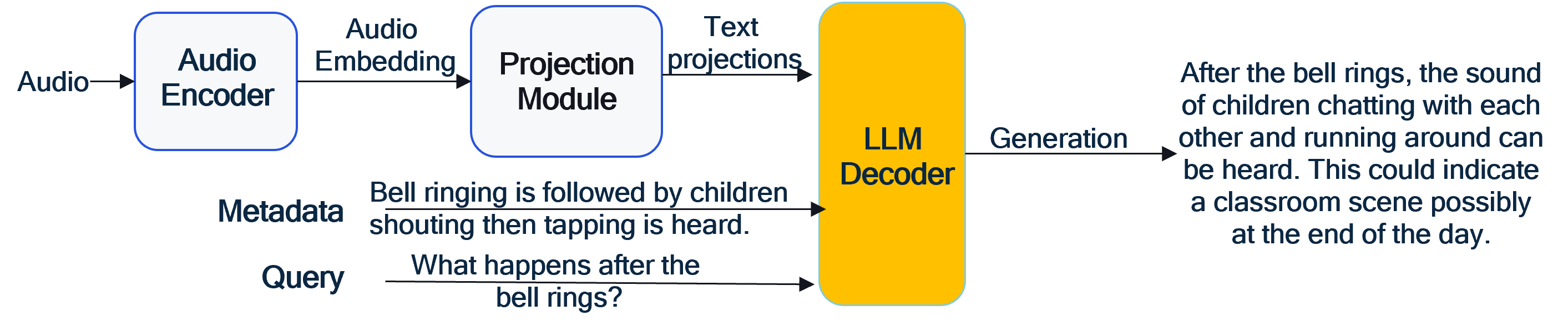}
  }
\end{minipage}
\caption{Our Proposed Framework for Audio Question Answering (AQA) model architecture}
\label{fig:AQA framework}
% \vspace{-7mm}
\end{figure*}

Multimodal Question Answering (MQA) involves generating relevant answers for multimedia inputs such as images, audio, and video, in response to user queries \cite{pan2024chain}. Following the success of large pretrained transformer models for MQA, audio-specialized question answering has gained traction. Audio Question Answering (AQA) is an audio-to-text task where, given an audio file and a question, the model produces an answer by analyzing the audio content.\\
 \textbf{Audio Question Answering: }Recent literature \cite{gong2023listen, ghosh2024gamalargeaudiolanguagemodel, tang2024salmonn, deshmukh2023pengi} in AQA develops end-to-end pretrained transformer-based architectures known as Large Audio Language Models (LALMs). Figure \ref{fig:AQA framework} provides a general framework for our AQA model architecture \cite{gong2023listen}. It comprises three components: an audio encoder, a projection module, and a text decoder. The Audio Spectrogram Transformer (AST) \cite{gong21b_interspeech} encodes the input audio clip into spectrogram feature representations. The projection module converts these audio feature representations into text-equivalent embeddings for the text decoder. The LLaMA model serves as the text LLM decoder, taking the converted audio feature embedding and the question as input. During training, we add metadata as an optional input that is generated by the proposed data augmentation in Section \ref{sec:TemporalAugmentation}. It helps provide extra guidance to the LLM decoder along with the text projections of the audio clip and aids in the overall audio-text representation learning. The GAMA model \cite{ghosh2024gamalargeaudiolanguagemodel} follows a similar architecture to LTU \cite{gong2023listen}, combining multiple types of audio features, including activations from multiple layers of AST, Audio Q-former, and a soft prompt that provides audio events information. In this paper, we intend to discuss a few problems and limitations that we discovered in the process of developing a LALM for commercial edge devices and explain our proposed techniques to overcome them. We chose LTU as the base model for our experiments over GAMA due to the ease of on-device implementation.\\
\textbf{Use Case Motivation: }Although LALMs excel at general audio understanding and have shown good overall performance in audio captioning, classification tasks, and open-ended reasoning tasks, there is a significant gap between LALM research and real-world product requirements. First, LALMs fine-tuned end-to-end with millions of audio-text samples do not capture fine-grained audio understanding well. Their performance isn't impressive on specialized reasoning tasks that require fine-grained understanding, such as temporal reasoning \cite{gong2023listen}. Audio temporal reasoning is the ability to understand the temporal context and relationship between events in the input media. Specialized audio temporal understanding has significant potential across various sectors for commercial adoption. In healthcare, it can be used for continuous monitoring and analysis of heartbeat and respiration over a period of time and provide useful analysis and recommendations to the user. In smart homes, it can enable advanced security monitoring with privacy protection by capturing and analyzing the sequence of events in live stream audio coming from sensors located in multiple areas. \cite{gong2023listen} explains that the lack of fine-grained understanding in LALMs might be due to performing temporal downsampling at the audio encoder-projection module juncture, which is a trade-off to save computational efficiency and limited training data for temporal analysis. In this paper, we address both these limitations while also keeping in mind the limitations in commercial LALMs, including low memory footprint, ease of on-device implementation, reliability, and minimal training compute. Due to the difficulty in procuring large amounts of pretraining data, expensive compute power, and time constraints, it is painstakingly difficult to retrain an LALM from scratch for improving a particular skill. On top of that, the large memory requirements of LALMs make it difficult to run them on low-compute edge devices. \\
\textbf{Existing Work on Temporal Reasoning in AQA: }In this paper, we focus on optimal training pipeline strategies to improve audio temporal understanding. Before the pre-trained transformers era, DAQA \cite{daqafayek} and ClothoAQA \cite{samuel_lipping_2022_6473207clothoaqa} proposed a synthetic rule based and crowd sourced audio temporal reasoning datasets respectively. \cite{ghosh2024compa} published an annotated benchmark to evaluate the audio encoders on compositional reasoning including order or occurrence of acoustic events. \cite{yuan2024tclaptemporalenhancedcontrastivelanguageaudio} discuss the limitations of CLAP encoder in capturing temporal information and propose a data augmentation strategy to improve the same.\\
\textbf{Motivation for Deploying AQA on Edge: }With the large memory requirements of LALMs scaling billions of parameters, the inference becomes expensive to run on cloud GPUs \cite{DESISLAVOV2023100857}. For commercial audio understanding use cases, such as smart home Internet of Things (IoT) and industrial IoT, where we can capture streams of audio from various sources such as machinery, front door, kitchen, etc., using a simple audio receiver, we need the AQA model on an always-on low-powered edge device for reasonable inference cost and preserving privacy by performing computation of audio on a self-contained edge CPU.\\
\textbf{Contributions: }To the best of our knowledge, we are the first to investigate the problem and limitations of audio temporal understanding in LALMs and address them from a commercialization perspective. Our contributions in this paper are as follows: First, we propose a data augmentation technique to reliably generate audio temporal question and answer pairs using GPT-4. Second, we show that fine-tuning the baseline checkpoint via curriculum learning helps improve the model’s temporal awareness and reasoning without losing its original performance. Finally, we implement the AQA to run on CPU locally for commercial edge applications.

% we overcome the fine-grained temporal understanding limitations of LALMs by introducing a small off-the-shelf Audio Context Detection (ACD) model during inference time as an input to the LALM. The ACD model predicts audio event tags along with their timestamps. 

\begin{figure*}[t!]
\centering
\begin{minipage}[b]{0.8\linewidth}
  \centering
  \centerline{
  \includegraphics[width=1.1\textwidth,scale=1.0]{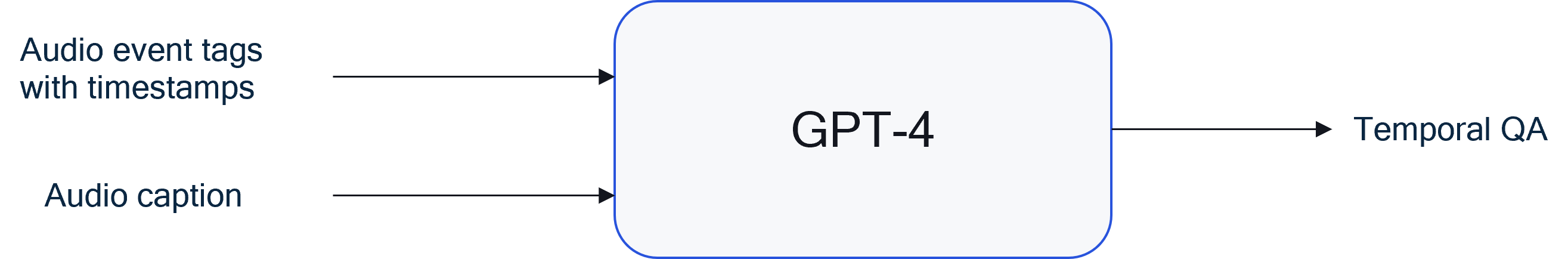}
  }
\end{minipage}
\caption{Data Generation for Temporal Reasoning using GPT-4}
\label{fig:Data gen framework}
% \vspace{-7mm}
\end{figure*}

 \begin{table*}[!ht]
    \centering
    \begin{tabular}{|p{2cm}|p{4cm}|p{3.8cm}|p{6cm}|}
    \hline
    Type & Metadata & Question & Answer \\ \hline
     Chronological order & The background of the audio is Music.The audio starts with Cough followed by Ratchet then Male singing. & What's the sequence of the audio events in the clip? & The chronological order of the sound events in the audio clip is as follows: First, a Cough is heard. This is immediately followed by a Ratchet sound. Finally, Male singing is heard. Hence, the sequence is cough, ratchet, and then male singing. \\ \hline
     Counting & The audio starts with Pig followed by Mechanisms then Pig followed by Mechanisms then Sound effect followed by Pig. & How many times does the Pig oink? & The Pig oink is heard 3 times with mechanisms sound occurring in between. This order forms a pattern where the pig sound reappears intermittently. \\ \hline
     Before/After & The background of the audio is Static.The audio starts with Male singing followed by Choir then Male singing followed by Choir. & What sequence of sound events follows the first choir singing? & Following the first choir singing, the sequence cycles back to male singing again, creating a repetitive pattern which is then followed by another choir performance.\\ \hline
     Duration & The background of the audio is Sawing.The audio starts with Speech. & What ambient sound persists throughout the audio clip? & The ambient sound that persists throughout the audio is sawing. This backdrop of sawing is mentioned as the continuous sound in the background.\\ \hline
     Temporal pattern detection & The background of the audio is Music.The audio starts with Child singing followed by Breathing then Child singing followed by Breathing then Child singing. & Which sound recurs after each instance of breathing? & The child's singing recurs after each instance of breathing. The pattern repeats multiple times in the audio.\\ \hline
    \end{tabular}
    \vspace{3mm}
    \caption{Types of temporal questions in TemporalQA}
    \label{Table: Data Augmentation Examples}
    % \vspace*{-5mm}
\end{table*}

\section{Methodology}
\label{sec:Methodology}
% \vspace{-2mm}
We divide our proposed methodology into two sections. First, we explain the data augmentation strategy for generating temporal reasoning data. Second, we discuss our temporal fine tuning strategy. 

% Third, we showcase the integration of Audio Context Detection (ACD) to improve fine-grained understanding during inference.
% \vspace{-3mm}
\subsection{Temporal Reasoning Data Augmentation}
\label{sec:TemporalAugmentation}
In order to improve the temporal reasoning capability of generalist LALMs, we developed a data augmentation technique that ensures the questions are intuitive to human temporal perception. We prompt GPT-4 \cite{openai2024gpt4technicalreport} with the audio clip’s ground truth, such as audio event labels, audio captions, and their metadata comprising timestamps of audio events. For Audioset-SL, we use audio clips containing fewer than seven types of foreground sound events with a total occurrence number of less than ten and consider sounds with a duration longer than nine seconds as background sound. We state the temporal information of the sound events in natural language and use that as part of the prompt for GPT-4 to generate question-answer pairs with rationale. In the prompt, we include few-shot paired examples of temporal reasoning tasks, including temporal relationships, chronological ordering, duration comparison, and occurrence counting. Table \ref{Table: Data Augmentation Examples} showcases the different types of metadata-question-answer pairs generated by our temporal data augmentation technique. We refer to the data generated by this pipeline as TemporalQA.

% \begin{table*}[!ht]
%     \centering
%     \begin{tabular}{|p{2cm}|p{3cm}|p{3cm}|p{1.3cm}|p{1.2cm}|p{0.9cm}|p{1.1cm}||p{1cm}|}
%     \hline
%     Question & Groundtruth Answer & LALM Response & BLEU-1 & ROUGE-L & SPIDER & FENSE & Open-Eval \\ \hline
%     \multirow{2}{2cm}{What sound event happens initially?} & \multirow{2}{3cm}{The audio clip begins with male speech. This clearly sets the tone and context right from the start, drawing listener's attention immediately to spoken words.} & The sound of a man speaking. &  0.005 & 0.054 & 0.0 & 0.495 & 0.7 \\ \cline{3-8}
%     & &  The audio starts with Male speech, which is a common way to begin a conversation or presentation in real-world scenarios. &   0.186 & 0.209 & 0.054 & 0.698 & 0.823 \\ \hline
%     \end{tabular}
%     \vspace{3mm}
%     \caption{Qualitative analysis of evaluation metrics on LALM responses.}
%     \label{Table:quali-openeval}
%     % \vspace*{-5mm}
% \end{table*}

\subsection{Temporal Finetuning via Curriculum Learning}
\label{sec:continual_finetuning}
In this section, we outline the training strategy employed to integrate temporal reasoning capabilities into a Large Audio Language Model (LALM) designed and finetuned for general audio understanding. To learn temporal reasoning skill on an already finetuned AQA model, we adopt a curriculum learning approach that merges TemporalQA with a few core finetuned Audio Question Answering (AQA) tasks, including audio classification and audio captioning. We conducted an empirical investigation to determine the optimal types of AQA tasks and the appropriate ratio of new skills (temporal reasoning) to existing skills. Based on our analysis and hyperparameter tuning, we observed that a 50:50 ratio of temporal reasoning to core AQA tasks—comprising audio event tagging, audio label classification, and audio captioning—combined with a learning rate ten times lower than that of the original finetuning, is optimal for learning temporal reasoning skills without significantly compromising the model’s original performance. We refer to our temporal finetuned model with and without metadata on LTU base as AQA+Temp-M and AQA+Temp, respectively.
\vspace{-2mm}
\begin{align*}
\small
% \vspace*{-2cm}
T_{\textrm{total}} = T_{\textrm{temporal}} + T_{\textrm{core AQA}},
 \tag{1}
 \label{eq:training data}
\end{align*}
Where T refers to training data and the + operation combines both operand datasets with a random shuffle. We also provide metadata of audio, such as audio events and background noise information, in natural language in the text prompt as guidance to mitigate the information bottleneck at the projection module.
 \begin{table*}[!ht]
    \centering
    \begin{tabular}{|p{1.8cm}|p{3.8cm}|p{4.8cm}|p{6cm}|}
    \hline
    Question & Groundtruth Answer & LTU & AQA-Temp (Ours)\\ \hline
     Does the whispering begin after the sound of the drum kit? & Yes, the whispering starts after the drum kit is heard. & Yes, it starts at 10.2 seconds. & Yes, the whispering starts after the sound of the drum kit. This suggests that the whispering is a separate event that occurs after the initial musical performance has begun. It's possible that the whispering is a response to the music or a way for the performers to engage with the audience in a more intimate setting. \\ \hline 
     What is the sequence of sound events following the first Speech? & Following the first instance of Speech, the sequence of sound events is as follows: Male speech, then another instance of Speech, followed by Male speech, and finally, Speech again. & The sequence includes a tap, followed by generic impact sounds and then more speech. This could suggest a casual conversation or activity involving objects being used or moved around in an outdoor setting & Following the initial speech, there are two instances of male speech. This suggests that the conversation continues with multiple participants, possibly a dialogue between two or more individuals. \\ \hline
    \end{tabular}
    \vspace{3mm}
    \caption{Demonstration examples showcasing temporal understanding after fine-tuning. LTU and LTU-Temp refers to the baseline and temporal reasoning finetuned checkpoints.}
    \label{Table:examples}
    % \vspace*{-5mm}
\end{table*}

\begin{table*}[!ht]
    \centering
    \begin{tabular}{|p{3.5cm}|p{1.2cm}|p{1.2cm}|p{1.2cm}|p{1.2cm}|p{1.2cm}|p{1.2cm}|p{1.2cm}|p{1.2cm}|}    
    \hline
    Model & \multicolumn{2}{|c|}{Clotho} & \multicolumn{2}{|c|}{AudioCaps} & \multicolumn{2}{|c|}{FSD} & \multicolumn{2}{|c|}{TemporalQA} \\
    \cline{2-9}    
    & SPIDER & FENSE  & SPIDER & FENSE  & SPIDER & FENSE  & SPIDER & FENSE  \\\hline 
    LTU & 0.19 & 0.56 & 0.31 & 0.67 & \textbf{0.08} & \textbf{0.47} & 0.27 & 0.57 \\ \hline
    GAMA & 0.04 & 0.41 & 0.09 & 0.55 & 0.05 & 0.42 & 0.22 & 0.65 \\ \hline \hline
    AQA+Temp (Ours) & 0.24 & 0.61 & 0.38 & 0.71 & 0.06 & 0.44 & 0.48 & 0.66  \\ \hline
    AQA+Temp-M (Ours) & \textbf{0.31} & \textbf{0.62} & \textbf{0.43} & \textbf{0.73} & 0.07 & 0.43 & \textbf{0.70} & \textbf{0.73} \\ \hline
    % LTU+Temp-M + ACD (proposed approach) & 
    \end{tabular}
    \vspace{3mm}
    \caption{Comparison of performance on LTU baseline with proposed finetuning on temporal reasoning. Temp refers to temporal finetuning and Temp-M refers to temporal finetuning with meta data information.}
    \label{Table:LTUresults}
    % \vspace*{-5mm}
\end{table*}

\section{Experiments}
% \vspace{-6mm}
\subsection{Datasets}
\label{sec:datasets}
We choose the LTU model \cite{gong2023listen} as our baseline. We adopt a similar training dataset accruing strategy to \cite{gong2023listen}. Our initial stages of curriculum learning focus on training the audio encoder and projection model with a combination of audio event classification public datasets, including Audioset, FSD50k, VGGSound, and Freesound, and audio captioning public datasets, such as Clotho and Audiocaps \cite{gong2023listen}. We use Audioset-strong labelled \cite{audiosetsl} and FSD50k datasets to synthetically generate 20k temporal reasoning data using the data augmentation strategy explained in Sec 2.1. TemporalQA has an 80:20 train-test split. We adopt the inference style of \cite{gong2023listen}, including the generation of audio descriptions for the FSD dataset. All audio clips are truncated to 10s to fit the audio encoder context window.

% \vspace{-2mm}
\subsection{Experiment Setup}
% \vspace{-1mm}
We train the AQA architecture from scratch with four-stage curriculum learning as described in \cite{gong2023listen}. For temporal reasoning fine-tuning, we perform model parallelism-based distributed training on 8 A100 GPUs for 2 epochs with a learning rate of 1e-4 and cross-entropy as the loss function. We found that a batch size of 24 and a micro-batch size of 1 work best for specializing the model further on a single task, as opposed to a batch size of 256 and a micro-batch size of 16 for fine-tuning from scratch. The low-rank adaptors (LoRA) hyperparameters alpha and r are set to 16 and 8, respectively. We set $\alpha_1$ and $\alpha_2$ to 1 while keeping $\alpha_3$ at 0 to provide equal weightage for answer and reason and for a fair comparison with conventional metrics.
% \vspace{-5mm}
\subsection{On-device Implementation}
To run the AQA model on CPU, we perform 16-bit and 8-bit post-training quantization as mentioned in \href{https://github.com/ggerganov/llama.cpp}{llama.cpp}. We implement the AQA architecture on top of the C++ implementation of LLaMA in the llama.cpp framework. First, we merge the LoRA weights into the LLaMA model of AQA+Temp and convert the checkpoint to gguf format. Second, we implement the audio encoder and projection module in \href{https://github.com/microsoft/onnxruntime}{onnxruntime} to combine their outputs with the LLaMA in C++. We perform the experiment to measure inference speed on 100 randomly sampled questions from our test set of AQA described in \ref{sec:datasets} and report the average.
\begin{table}[htbp]
    % \tiny
    \scriptsize
    \centering
    \begin{tabular}{ccc}
    \toprule % replaced all \hline commands with rules from the booktabs package
\textup{Model Name} & Size & Accuracy(\%) \\
        \midrule
    Random Guess & - & 26.72\\
    Most Frequent Choice & - & 27.02 \\
     Human (test-mini) & - & 86.31 \\ \hline
     Pengi & 323 M & 6.1 \\
     Audio Flamingo Chat & 2.2B & 23.42 \\ \hline
    M2UGen & 7B & 3.6 \\
    LTU & 7B & 22.52 \\ 
    LTU AS & 7B & 23.35 \\
    MusiLingo & 7B & 23.12 \\
    MuLLaMA & 7B & 40.84 \\
    GAMA & 7B & 41.44 \\
    GAMA-IT & 7B & \textbf{43.24} \\ \hline
    Qwen-Audio-Chat & 8.4B & 55.25 \\
    Qwen2-Audio & 8.4B & 7.5 \\
    Qwen2-Audio-Instruct & 8.4B & 54.95 \\
    SALAMONN & 13B & 41 \\
    Gemini Pro v1.5 & - & 56.75 \\
    GPT4o + weak cap. & - & 39.33 \\
    GPT40 + strong cap. & - & 57.35 \\
    Llama-3-Instruct + weak cap. & 8B & 34.23 \\
    Llama-3-Instruct + strong cap. & 8B & 50.75 \\ \hline
    % LTU + ACD (Ours) & 7B & 27.63 \\
    AQA+Temp (Ours) & 7B & 28.83 \\
    AQA+Temp-M (Ours) & 7B & \textbf{32.73} \\
    
         \bottomrule
    \end{tabular}
    \caption{Results on MMAU Test-Mini Sound Split}
    \label{tab:MMAUbenchmark}
\end{table}

\begin{table*}[htbp]
    \centering
    \begin{tabular}{ccccc}
    \toprule % replaced all \hline commands with rules from the booktabs package
\textup{Float Precision (bits)} & Model Size (GiB) & Load Time (ms) & Prompt Eval Rate (TPS) & Eval Rate (TPS)\\
        \midrule
    16 & 12.55 & 10925.84 & 6.95 & 7.35\\
    8 & 6.67 & 2690.79 & 13.57 & 13.16 \\
     4 & 3.56 & 1395.71 & 15.79 & 19.64\\
         \bottomrule
    \end{tabular}
    \caption{Comparison of inference speed for AQA across different floating point precision and devices. TPS refers to Tokens per second}
    \label{tab:cpuspeed}
\end{table*}

\begin{figure}[t!]
\begin{minipage}[b]{1.2\linewidth}
  \centering
  \centerline{
  \includegraphics[width=1.0\textwidth,scale=0.8]{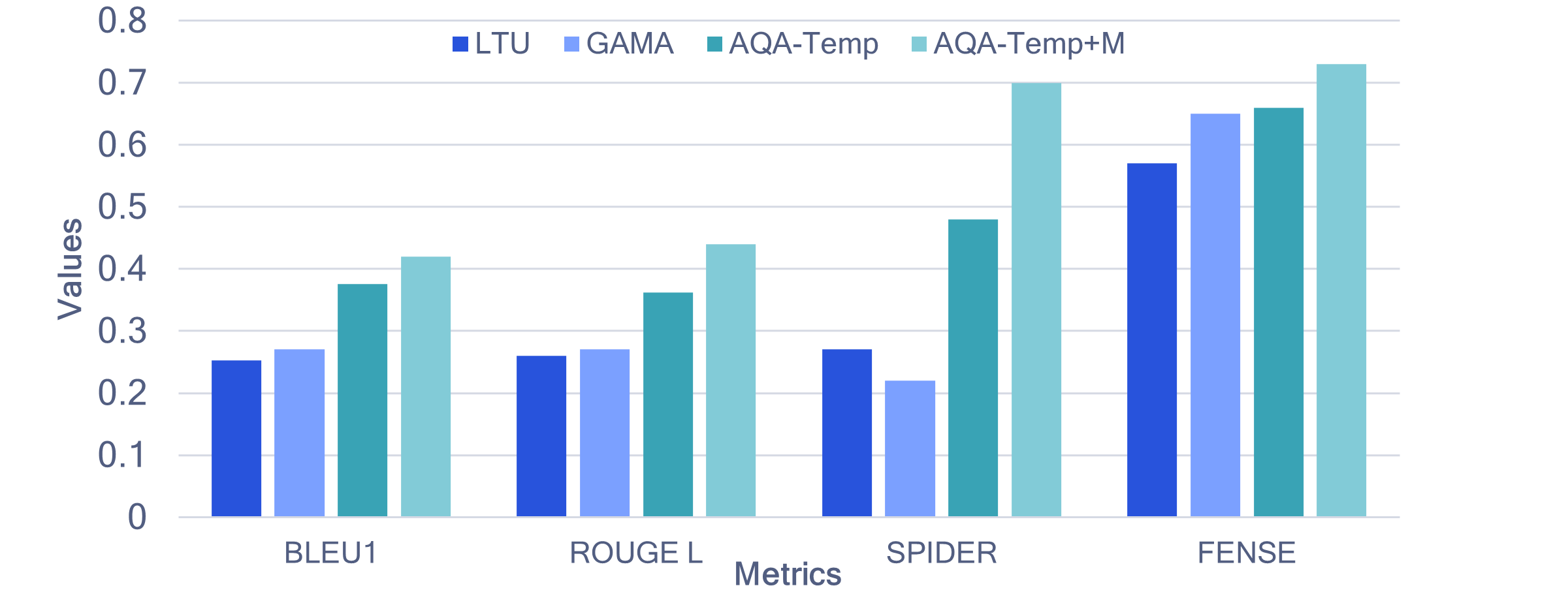}
  }
\end{minipage}
\caption{Barplot of LTU and GAMA baseline and temporal finetuned checkpoints for temporal dataset.}
% \vspace{-5mm}
\label{fig:LTUbarplot}
\end{figure}

% \begin{figure}[t!]
% \begin{minipage}[b]{1.0\linewidth}
%   \centering
%   \centerline{
%   \includegraphics[width=1.0\textwidth,scale=1.0]{latex/gama_barplot_latest.png}
%   }
% \end{minipage}
% \caption{Barplot of GAMA baseline and temporal finetuned checkpoint for temporal dataset.}
% \label{fig:GAMAbarplot}
% \end{figure}

% \begin{table*}[!ht]
%     \centering
%     \begin{tabular}{|p{2.3cm}|p{1.2cm}|p{1.2cm}|p{1.2cm}|p{1.2cm}|p{1.2cm}|p{1.2cm}|p{1.2cm}|p{1.2cm}|}    
%     \hline
%     Model & \multicolumn{2}{|c|}{Clotho} & \multicolumn{2}{|c|}{AudioCaps} & \multicolumn{2}{|c|}{FSD} & \multicolumn{2}{|c|}{Temporal} \\
%     \cline{2-9}    
%     & SPIDER & SentBert & SPIDER & SentBert & SPIDER & SentBert  & SPIDER & SentBert \\\hline 
%     GAMA & 0.04 & 0.86 & 0.09 & 0.87 & 0.05 & 0.82 & 0.22 & 0.89\\ \hline
%     GAMA+Temp & 0.21 & 0.91 & 0.34 & 0.92 & 0.06 & 0.83 & 0.46 & 0.90 \\ \hline

%     \end{tabular}
%     \vspace{3mm}
%     \caption{Comparison of performance on GAMA baseline with proposed temporal finetuning on temporal reasoning. Temp refers to temporal finetuning.}
%     \label{Table:GAMAresults}
%     \vspace*{-7mm}
% \end{table*}
\section{Results}
\vspace{-2mm}
\subsection{Quantitative Analysis of Temporal Finetuning:}
Table \ref{Table:LTUresults} shows the performance of the proposed temporal fine-tuning for temporal reasoning with LTU as the base model. For a fair evaluation, during inference, we do not provide metadata to the models. After temporal fine-tuning, there is a considerable increase in all the metrics across datasets except for FSD. This could be due to the fact that FSD provides minimal ground truth information, being a classification dataset. The significant improvement of Spider and FENSE metrics for AQA+Temp-M over LTU shows that we can offset the information bottleneck at the projection layer to some extent with extra textual guidance. It is notable that our AQA+Temp and AQA+Temp-M models performs better than the GAMA baseline, which has a sophisticated audio encoding. This emphasizes the need for good data augmentation in addition to architectural improvements. From the reasonable improvement in scores across all the datasets of AQA+Temp-M compared to AQA+Temp, we infer that providing metadata during training helps in better detection of audio events and improved audio-text representation mapping. In Fig \ref{fig:LTUbarplot}, our proposed models show consistent improvements over the baseline, indicating the effectiveness of temporal finetuning. \\
Table \ref{tab:MMAUbenchmark} presents the performance of various models on the MMAU Test-Mini Sound split benchmark \cite{sakshi2024mmaumassivemultitaskaudio}. Based on our organization's guidelines, we use the test-mini instead of the full test set as the latter requires us to upload our model's generations to the MMAU webpage. Our proposed method, AQA+Temp-M, performs better than the baseline LTU by a significant margin of 10.21. This shows the efficacy of our proposed data augmentation and temporal finetuning. Hence, the proposed method improves temporal reasoning in the baseline LALM while maintaining previously learned skills, as illustrated quantitatively in Table \ref{Table:LTUresults} and \ref{tab:MMAUbenchmark}.
\subsection{Qualitative Analysis of Temporal Finetuning:}
From Table \ref{Table:examples}, it is evident that temporal finetuning with temporal reasoning data augmentation, as described in Section \ref{sec:continual_finetuning}, results in the generation of rationale with temporal commonsense knowledge compared to the baseline. In the first example, although the baseline's answer is correct, the reasoning is wrong since the model is only provided with 10 seconds of audio clip content. In the second example, the baseline model states incorrect audio events—tap and generic impact sounds—and continues to use them in the rationale. On the other hand, the AQA+Temp generates the correct temporal answer along with a plausible explanation as rationale. This illustrates a qualitative improvement in our proposed method's answer generation over the baseline.

\subsection{Insight on On-device AQA Inference: }
Table \ref{tab:cpuspeed} presents the model loading time and inference speed of AQA for different floating point precisions. The load time denotes the time taken to load the model into the CPU. Prompt Eval Rate measures the number of user query prompt tokens encoded relative to the time taken for performing audio and prompt encoding. Eval rate refers to the time taken to generate the response. User prompts should usually be encoded quicker than the response generation because user prompts can be encoded as a batch of tokens while a response is generated auto-regressively, word by word. Despite this, for the 4-bit and 16-bit models, we see a lower Prompt Eval Rate than Eval Rate. This could be due to the audio encoding overhead, which needs to be kept in mind for improving overall inference latency.

% \vspace{-6mm}
\section{Conclusion}
% \vspace{-3mm}
In this work, we proposed a novel data augmentation strategy to generate temporal reasoning QA pairs using LLMs. Next, we finetuned a SOTA AQA model on the generated temporal reasoning data and showcased quantitative improvements across evaluation metrics. Finally, we showcased our implementation of the AQA model on-device and studied its performance. In the future, we plan to investigate quantization-aware fine-tuning techniques and study the generation quality vs. quantization tradeoff. We plan to introduce an evaluation metric that can appropriately select the facts from the answer and compare them against the ground truth. We can use the metric as a loss term during fine-tuning of the AQA model to prioritize the learning of specialized skills reliably.

% \subsection{Appendices}

% Use \verb|\appendix| before any appendix section to switch the section numbering over to letters. See Appendix~\ref{sec:appendix} for an example.

% \section*{Acknowledgments}

\bibliography{acl_latex}

\appendix

\section{OnDevice Graphical User Interface Examples}
\label{sec:appendix_ondevice}
\begin{figure*}[t!]
\centering
\begin{minipage}[b]{1\linewidth}
  \centering
  \centerline{
  \includegraphics[width=1.1\textwidth,scale=1.0]{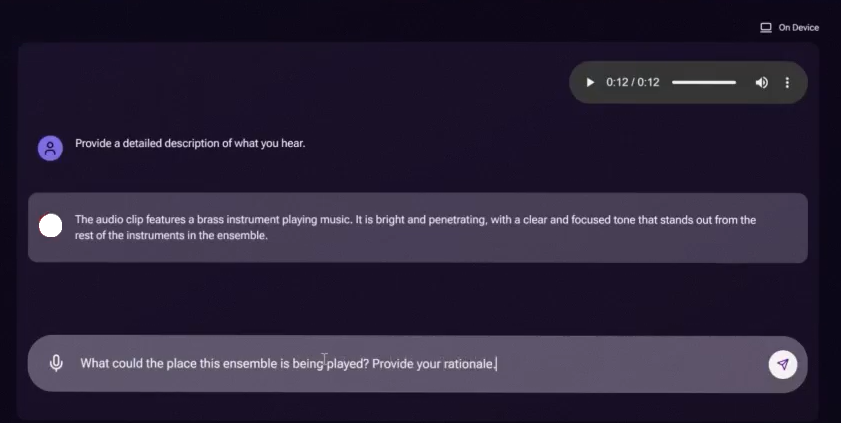}
  }
\end{minipage}
\caption{I: Graphical User Interface (GUI) of Audio Question Answering on ARM CPU.}
\label{fig:AQA GUI Demo part1}
% \vspace{-7mm}
\end{figure*}

\begin{figure*}[t!]
\centering
\begin{minipage}[b]{1\linewidth}
  \centering
  \centerline{
  \includegraphics[width=1.1\textwidth,scale=1.0]{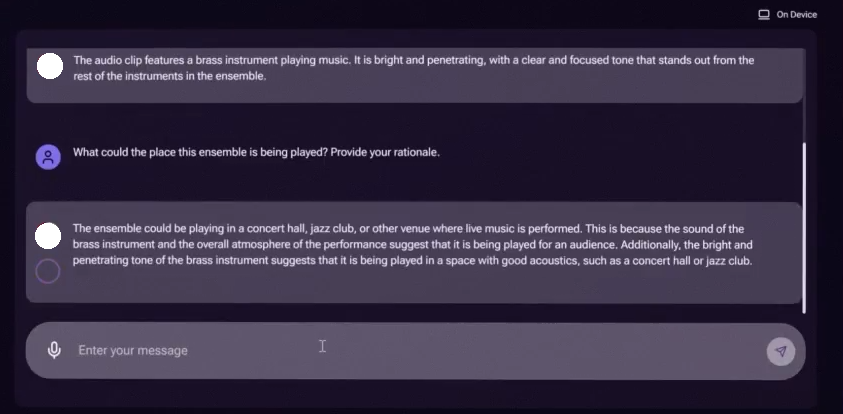}
  }
\end{minipage}
\caption{II: Graphical User Interface (GUI) of Audio Question Answering on ARM CPU.}
\label{fig:AQA GUI Demo part2}
\end{figure*}

Figure \ref{fig:AQA GUI Demo part1} and \ref{fig:AQA GUI Demo part2} shows the GUI and an example sample for AQA running on edge CPU.

\section{System Prompts}
\label{sec:System Prompt}
The System Prompts used for generating temporal question answering data and for on-device inference are shown in Table \ref{Table:systempprompt}.

\section{Sample Conversation with AQA}
\begin{figure*}[t!]
\centering
\begin{minipage}[b]{0.7\linewidth}
  \centering
  \centerline{
  \includegraphics[width=1.1\textwidth,scale=1.0]{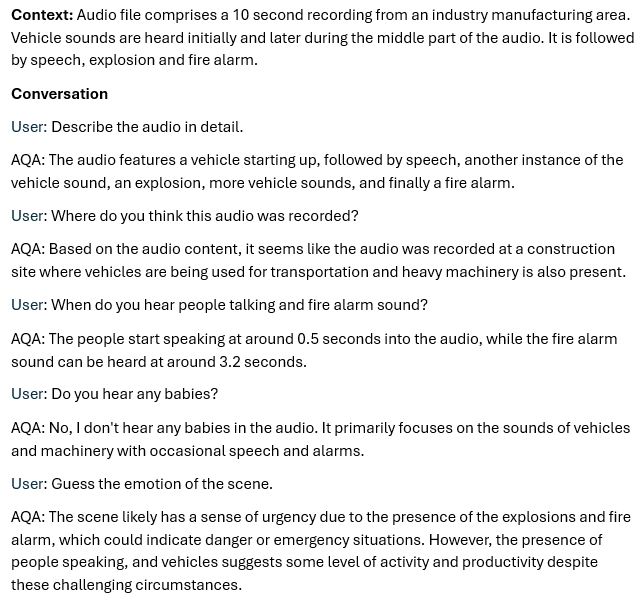}
  }
\end{minipage}
\caption{Example of a conversation with AQA-Temp proposed model}
\label{fig:SampleConversation}
\end{figure*}

Figure \ref{fig:SampleConversation} shows a sample conversation with AQA on an audio file recorded in an industrial setting.

\section{Device Specifications for the on-device demo}
\label{sec:DeviceSpec}
The Device has an ARM-based Snapdragon(R) X Elite processor with 32.0 GB RAM (31.6 GB usable). The CPU has 3.42 GHz clock speed operating on a 64-bit operating system.

 \begin{table*}[!ht]
    \centering
    \begin{tabular}{|p{3cm}|p{13cm}|}
    \hline
    Stage & System Prompt\\ \hline
     Temporal Data Generation & Generate 5 questions and answer pairs along with metadata from the following information about the audio. The questions are used for temporal audio question answering task. Assume the audio description and audio event time information as the audio file itself. Do not ask questions whose answers are not present in the description. Write the answers in a more explanatory and human friendly manner. You can add some common senses or facts whenever it is possible along with the answer. Format each question in a single line as a JSON dictionary with keys - "id", "question", "answer", "metadata". Some examples of questions you could ask are :  What sound events occurs first?  What sound comes after the male speech at the beginning? (if male speech is present in the description) What event happens before the engine running sound?  Which event occurs towards the end ?    Is the door bell sound after the dog barking? Answer true or false and provide your reasoning steps.     Can you hear footsteps before the baby cries? Answer true or false and provide your reasoning steps.  What is the chronological order of the sound events?   What is the background sound if there's any? Please generate diverse questions with paraphrasing.\\ \hline 
     AQA On-device Inference & A chat between a curious user and an audio question answering artificial intelligence assistant. The assistant gives helpful, detailed, and polite answers to the user's questions. You are given an audio clip and a question from the user. Do not generate false audio events or hallucinations that are not there in the audio clip. Do not contradict yourself without proper evidence.\\ \hline

    \end{tabular}
    \vspace{3mm}
    \caption{Systemp Prompts for Temporal Data Generation and AQA On-device Inference.}
    \label{Table:systempprompt}
    % \vspace*{-5mm}
\end{table*}

\end{document}